\documentclass[12pt]{article}
\usepackage{graphicx}

\newcommand{\imp}{\mbox{\boldmath $p$}}
\newcommand{\Ebf}{\mbox{\boldmath $E$}}
\newcommand{\Pbold}{\mbox{\boldmath $P$}}
\newcommand{\Hbf}{\mbox{\boldmath $H$}}
\newcommand{\rbf}{\mbox{\boldmath $r$}}
\newcommand{\zbf}{\mbox{\boldmath $z$}}
\newcommand{\nbf}{\mbox{\boldmath $n$}}
\newcommand{\phibf}{\mbox{\boldmath $\phi$}}
\newcommand{\Phibf}{\mbox{\boldmath $\Phi$}}
\newcommand{\nablabf}{\mbox{\boldmath $\nabla$}}
\newcommand{\BB}{{\rm BB}}
\newcommand{\BL}{{\rm BL}}
\newcommand{\pa}{\partial}

\newcommand{\Fcal}{{\cal F}}
\newcommand{\Ecal}{{\cal E}}
\newcommand{\grm}{{\rm g}}
\newcommand{\Rerm}{{\rm Re}}

\newcommand{\la}{\lambda}

\newcommand{\al}{\alpha}
\newcommand{\ze}{\zeta}
\newcommand{\eps}{\varepsilon}

\newcommand{\bb}{\begin{equation}}
\newcommand{\ee}{\end{equation}}
\newcommand{\bega}{\begin{eqnarray}}
\newcommand{\ega}{\end{eqnarray}}
\newcommand{\begae}{\begin{eqnarray*}}
\newcommand{\egae}{\end{eqnarray*}}

\newcommand{\h}{\hspace*{4ex}}
\newcommand{\dis}{\displaystyle}

\newcommand{\om}{\omega}

\newcommand{\ove}{\overline}

\newcommand{\cent}{\centerline}
\newcommand{\vs}{\vspace*}

\begin{document}

\baselineskip 0.6cm

\begin{center}

{\large {\bf On Localized ``X-shaped'' Superluminal Solutions \\
to Maxwell Equations}$^{\: (\dag)}$}
\footnotetext{$^{\: (\dag)}$  Work partially supported by INFN, MURST and
CNR (Italy), and by CAPES, CNPq (Brazil). \ {\bf This paper first appeared in
preliminary form as Report INFN/FM--96/01 (I.N.F.N.; Frascati, 1996),
its number in the electronic LANL Archives being \# physics/9610012; \
Oct.15, 1996: cf. ref.[24].}}

\end{center}

\vs{5mm}

\cent{ Erasmo Recami }

\vs{0.5 cm}

\cent{{\em Facolt\`{a} di Ingegneria, Universit\`{a} Statale di Bergamo, Dalmine
(BG), Italy;}}
\cent{{\em INFN---Sezione di Milano, Milan, Italy; \ and}}
\cent{{\em DMO--FEEC and CCS, State University of Campinas,
Campinas, S.P., Brazil.}}

\vs{1. cm}

\centerline{\small{This is a revised version of a paper submitted on March 1, 1996,}}
\centerline{\small{to appear in Physica A with the original submission date (03.01.1966)}}

\vs{1. cm}

{\bf Abstract  \ --} \ In this paper we extend for the case of Maxwell
equations the ``X-shaped" solutions previously found in the case of scalar
(e.g., acoustic) wave equations. Such solutions are localized in theory:
i.e., diffraction-free and particle-like (wavelets), in that
they maintain their shape as they propagate. \ In the electromagnetic case
they are particularly interesting, since they are expected to be
Superluminal. \ We address also the problem of their practical, approximate
production by finite (dynamic) radiators. \ Finally, we discuss the
appearance of the X-shaped solutions from the purely geometric point of
view of the Special Relativity theory.\\

PACS nos.: \ 03.50.De ; \ \ 41.20.Jb ; \ \ 03.30.+p ; \ \ 03.40.Kf ; \ \
14.80.-j \hfill\break


\newpage

{\bf I. -- INTRODUCTION}\\

\h Starting with the 1915 pioneering work by H.Bateman[1], it became slowly
known that all the relativistic homogeneous wave equations ---in a
general sense: scalar, electromagnetic and spinorial--- admit also solutions
with group velocities slower than the ordinary wave velocity in
the considered medium.  More recently, solutions had been constructed for
those homogeneous wave equations with group velocities even higher than the
ordinary wave velocity in the medium[2].

\h In the case of acoustic waves, for instance, the existence has been shown
of ``sub-sonic" and ``Super-sonic" solutions in some well-known
papers[3-6]. We may expect such solutions to exist also for seismic
wave equations (and perhaps in the case of gravitational waves too).

\h Here we shall fix our attention on Maxwell equations, which is the most
intriguing case since their new solutions will be subluminal and Superluminal,
respectively (i.e., with group velocities in vacuum lower and higher than
$c$, respectively). \ In ref.[2], e.g., one Superluminal solution was found
just by applying a Superluminal Lorentz ``transformation"[7,8].

\h Only in the last few years the ``subluminal" and ``Superluminal"
solutions of wave equations have been systematically investigated, in
particular by Donnelly and Ziolkowski[9] (see also refs.[10]).

\h Most of the attention, till now, has been paid however to the fact that
some ---both among the subluminal and among the Superluminal new solutions---
are {\em localized} (wavelet-type) beams[11], which theoretically can
propagate to an infinite distance without changing their wave shape; in
other words they are ``undistorted progressive waves", to use the Courant
and Hilbert's terminology[12]. \ Localized beams were discovered as early as
1941 by Stratton[13] and rediscovered by Durnin[14] in 1987. These beams
have a priori an infinite depth of field. Durnin termed these beams
``nondiffracting beams", or ``diffraction-free beams", while a more realistic
name would be that of ``limited diffraction" (or, rather,
``limited diffraction") beams since in all the {\em practically realizable}
situations they will diffract  eventually[15-17].
Durnin's beams are also called Bessel beams because their transverse beam
profile is a Bessel function[14]. Bessel beams are monochromatic and are
obtained by treating as usual the (scalar) amplitude related to one
transverse component only of either the electric or the magnetic field of
light as a solution to the scalar wave equation; in a sense, they are the
cylindrical counterpart of the plane-wave solutions.  It is interesting
that the Bessel beam essentially maintains its propagation--invariance
property even in its physically realizable approximate versions[14]. \
Due to their large ``depth of field" with a zero diffraction angle,
limited-diffraction beams are having applications in non-destructive
evaluation of materials[18], Doppler velocity estimation, tissue
characterization, and particularly medical imaging[3,16,17]. \ Besides
in acoustics[3,16-18], they can have applications in
electromagnetism[19-21,23-25] and in optics[14,22,26,27].

\h But localized, non-diffractive solutions to homogeneous wave equations can
play an important theoretical r\^{o}le in non-relativistic and relativistic
Quantum Mechanics for a suitable description of elementary particles[28]
(usually described by unsuitable wave-packets that {\em spread} even in
vacuum).\\

\

{\bf II. -- LOCALIZED SOLUTIONS TO FREE MAXWELL EQUATIONS}\\

\h Localized beams have been studied extensively in recent years
in both acoustics[29,30] and optics[22,27].

\h Recently, a new family of localized beams has been discovered[4]. These
beams have been called ``X waves" because they are X-shaped in a plane
($r, z$) passing through the propagation axis[5,31,32]. The X-shaped
waves are different from the Bessel beams because they contain multiple
frequencies, but possess the extremely important characteristic of being
{\em non-diffractive} in isotropic-homogeneous media or free space[4,5,23]. \
Let us recall that Bessel beams are ``localized" at a single frequency,
but become diffractive for multiple frequencies because the phase velocity
of each frequency component of theirs is different[33].

\h Even more important, for us, is the fact that the X-shaped waves are
Superluminal[5,6,9,23], i.e., propagate rigidly with Superluminal speed.

\h In cylindrical coordinates, localized beams propagating
along the $z$ axis can be written in the following form

\hfill{$
\Phi (r, \phi, z - c_1 t) \ ,
$\hfill} (1)

\

where: $r,\phi,z$ and $t$ represent the radial distance, polar angle, axial
distance, and time, respectively; $\Phi$ represents the Hertz potential (or,
in other cases, the acoustic pressure, or the velocity potential);
$z-c_{1}t$ is a propagation term, and $c_{1}$ is the velocity of the beam.
Because the variables, $z$ and $t$, appear only in the propagation term
in eq.(1), localized beams are only a function of $r$ and $\phi$ if
$z-c_{1}t = {\rm const.}$; that is to say, travelling with the beam at the
speed of $c_{1}$, one sees a constant beam pattern. This is different
from conventional focused beams[34] and from the localized waves studied by
Brittingham[19] and other investigators[20,21,35].

\h In Section 3, we shall extend the theory of the X-shaped beams to
electromagnetic waves, i.e., we shall consider X-shaped wave solutions to
the free Maxwell equations.\\

\h Let us start by recalling the well-known fact that, even if Maxwell's are
vector equations, in various
cases such as optics and microwaves, they can be simplified, i.e., only
the scalar amplitude of one transverse component of either the electric
field $\Ebf$ or the magnetic field $\Hbf$ strength is considered, and any
other components of interest are treated independently in a similar fashion
(treating light and microwaves as a scalar phenomenon). This is approximately
true ---for ordinary experimental setups---  under the following
conditions[36]: (i) the diffracting aperture must be large compared with a
wavelength, and (ii) the diffracted fields must not be observed too close
to the aperture. In this case, localized beams developed in acoustics[3,16-21]
can be directly applied to electromagnetism because they share the same
wave equation (this was verified even experimentally in optics by Durnin et
al. for Bessel beams[14]).

\h Another way to solve the Maxwell equations is to use the (magnetic) Hertz
{\em vector} potential \ $\Phibf = \Phi {\hat{\nbf}}$, where ${\hat{\nbf}}$
is a unit vector (this implies that the electromagnetic wave given by the
Maxwell equation $\nablabf {\cdot} \Ebf = 0$ is a TE (``Transverse Electric
field") polarization wave that is perpendicular to ${\hat{\nbf}}$; for TM
(``Transverse Magnetic field") polarization, the procedure is similar). \
This approach is rigorous, as opposed to the scalar method above. But one
easily gets (cf., e.g., ref.[24]) expressions for  $\Ebf$ and $\Hbf$ in
terms of $\Phi$, where the Hertz vector potential is still a solution of
the scalar wave equation. \ Of course, not all the solutions of the scalar
wave equation are localized.  However, if $\Phi$ is a localized solution,
then also the solution of the Maxwell equations is localized (since the
derivatives with respect to the variables do not change[16] the propagation
term $z-c_{1}t$). \ Because of this, numerous limited diffraction
(relativistic) {\em electromagnetic waves} can be obtained from the scalar
localized (non-relativistic) beams studied in acoustics.

\h Actually, families of generalized solutions of the equation

\

\hfill{$
\left( \nabla^2 - \dis{{1 \over c^2} {{\pa^2}  \over {\pa t^2}}} \right) \Phi = 0 \ ,
$\hfill} (2)

\

were discovered recently in the acoustic case[4]. One of the families of
solutions is given by[24]:

\hfill{$ \Phi_\ze (s) \ = \ \int_0^\infty T(k) \; \left[ {1 \over
{2\pi}} \int_{-\pi}^\pi A(\theta) \; f(s) \; d\theta \right] \; dk
\ , $\hfill} (3)

where

\hfill{$ s \ \equiv \ \al_0(k,\ze) r \cos (\phi-\theta) + b(k,\ze)
[z {\pm} c_1(k,\ze)
 \, t] \ ,
$\hfill} (3a)

and where

\hfill{$
c_1(k,\ze) \ \equiv \ c \sqrt{1 + [\al_0(k,\ze) / b(k,\ze)]^2} \ .
$\hfill} (3b)

\

The quantity $T(k)$ is any complex function (well behaved) of $k$
and can include the temporal frequency transfer function of a
practical {\em electromagnetic} antenna (or acoustic transducer);
\ $A(\theta)$ is any complex function (well behaved) of $\theta$
and represents a weighting function of the integration with
respect to $\theta$; \ quantity $f(s)$ is any complex function
(well behaved) of $s$; \ quantities $\al_0(k,\ze)$  and $b(k,\ze)$
are any complex function of $k$ and $\ze$; while $c$ is the {\em
speed of light} (or of sound) entering eq.(2), and $k$, $\ze$ are
variables that are independent of the spatial position $\rbf = (r
\cos \phi, r \sin \phi, z)$ and of time $t$. \ At last, $\ze$ is
the Axicon angle [4,37], that we confine in the range \ $0 < \ze <
\pi/2$.

\h If $c_1(k,\ze)$ in eq.(3b) is real, then ``${\pm}$" in eq.(3a)
represent forward and backward propagating waves, respectively (in the
following analysis, we consider only the forward propagating waves and
all results will be the same for the backward propagating waves).
Furthermore, $\Phi_\ze(s)$ will represent a family of localized waves
if $c_1(k,\ze)$ is independent of $k$ (containing, that is, the same
propagation terms $z-c_1(\ze) \; t$ for all frequency components $k$).
 \ It must be noticed that $\Phi_\ze(s)$ in eq.(3) is very general. It
contains some of the localized solutions known previously, such as
the plane wave and Durnin's localized beams, in addition to a quantity of
new beams.\\

{\bf III. -- ELECTROMAGNETIC X-SHAPED WAVES}\\

\h We shall now find that the localized X-shaped waves discovered in the
scalar case[4] exist also in electromagnetism, i.e., hold also as solutions
to Maxwell equations, by use of the Hertz potential.  Let us recall that,
in terms of $\Phi$, when ${\hat{\nbf}}$ is chosen in the $z$-direction, the
quantities $\Ebf$ and $\Hbf$ read[24]

\

\hfill{$
\Ebf = -\mu_0 \dis{{1 \over r} {{\pa^2 \Phi} \over {\pa t \pa \phi}}} {\hat{\rbf}} +
\mu_0 \dis{{{\pa^2 \Phi} \over {\pa \phi \pa r}}} {\hat{\phi}}
$\hfill} (3')

and

\hfill{$
\Hbf =  \dis{{{\pa^2 \Phi} \over {\pa r \pa z}}} {\hat{\rbf}} + \dis{{1 \over r}
{{\pa^2 \Phi} \over {\pa \phi \pa z}}} {\hat{\phi}} + \left( \dis{{{\pa^2 \Phi} \over
{\pa z^2}} - {1 \over c^2} {{\pa^2 \Phi} \over {\pa t^2}}} \right) {\hat{z}} \ ,
$\hfill} (3'')

respectively.\\

{\em A. -- X-Shaped Wave Solutions}

\h Let us consider eq.(3). \ When $T(k) = B(k) \exp[-a_0 k]$; \
$A(\theta) = i^n \exp[in\theta]$; \ $\al_0(k,\ze) = -ik
\sin{\ze}$; \ $b(k,\ze) = ik \cos{\ze}$ and $f(s) = \exp[s]$, we
obtain an $n\/$th-order scalar localized ``X-wave"[24] that has an
X-like shape in a plane ($r,z$) passing through the propagation
axis~$z$:

\hfill{$
\Phi_{X_n} (r, \phi, z - c_1 t) \ = \hfill\break
 \hfill = \ e^{in\phi} \; \int_0^\infty \; B(k) J_n(kr \sin \ze) e^{-k [a_0 - i
 \cos \ze (z - c_1 t)]} \; dk \ , \ \ \ \ \ (n = 0, 1, 2, \ldots) \ ,
$\hfill} (4)

\

where $B(k)$ is any well-behaved function of $k$ (representing as before the
transfer function of an electromagnetic antenna); the quantity  $J_n({\cdot})$
is the $n\/$th-order {\em Bessel function} of the first kind; \ $c_1 = c /
\cos{\ze}$; \ $k = \om / c$; \ $\om$ is the angular frequency; while $a_0 > 0$
\ and \ $0 < \ze < \pi/2$ are constant.

\h It can be immediately noticed that $c_1$ is larger than the light speed
in the medium; i.e., in vacuum, is {\em larger} than $c$. \ Let us recall
that one can get the group velocity $v_\grm$ by the stationary phase
method[38] (provided the considered wave-packet presents a clear bump), i.e.,
by equating to zero the partial derivative with respect to $k$ of the unitary
phase factor entering eq.(4): \ $\pa [k \cos\ze - c_1 k \cos \ze t] / \pa k \
= \ 0$, \ which yields \ $z - c_1 t = 0$ \ and therefore \ $v_\grm = c_1$.
Notice that for each component it is $v_\grm = d\Ecal / dk$, and $v_\grm$
depends only on the relation $\Ecal = \Ecal(k)$ (the quantity $\Ecal$ being
the energy).

\h If $B(k) = a_0$, from eq.(4) we get the $n\/$th-order broadband[4,24]
X-wave

\

\hfill{$
\Phi_{X \BB_n} (r, \phi, z - c_1 t) \ = \
\dis{{{a_0 (r \sin \ze)^n e^{in\phi}} \over {\sqrt{M} (\tau + \sqrt{M})^n}}} \ ,
 \ \ \ (n = 0, 1, 2, \ldots) \ ,
$\hfill} (4')

\

where the subscript ``BB" means ``broadband"; \ $M = (r \sin \ze)^2 + r^2$,
 \ and \ $\tau = a_0 - i \cos \ze (z - c_1 t)$.

\h If $B(k)$ is a band-limited function[4,24], we obtain a $n\/$th-order
band-limited X-wave that is a convolution of functions $\Fcal^{-1} [B({\om
\over c})] / a_0$ \ and \ $\Phi_{X \BB_n} (r, \phi, z - c_1 t)$ \ with respect
to time~$t$

\

\hfill{$
\Phi_{X \BL_n} (r, \phi, z - c_1 t) \ = \
 \dis{{1 \over a_0}} \; {\Fcal}^{-1} \left[ B{\dis{({\om \over c})}}
 \right] \; * \; \Phi_{X \BB_n} \ , \ \ \ (n = 0, 1, 2, \ldots) \ ,
$\hfill} (4'')

\

where ${\Fcal}^{-1}$ represents the inverse Fourier transform, the star ``*"
denotes the convolution, and subscript ``BL" means ``band-limited".

\h By eq.(4'), one obtains finally the $n\/$th-order broadband
{\em electromagnetic X-shaped wave\/}[24]:

\hfill{$
(\Ebf_{X \BB_n})_r  \ = \ -\dis{{{n \mu_0 c} \over {rM}}} \; \left( \tau + n
 \sqrt{M} \right) \; \Phi_{X \BB_n} \ ,
$\hfill} (5a)

\

\hfill{$
(\Ebf_{X \BB_n})_\phi \ = \ \dis{{{i \mu_0 c} \over {rM}}} \; \Phi_{X \BB_n} {\cdot} \hfill\break
\hfill {\cdot} \left\{ \dis{{{3 \tau + 2n \sqrt{M}} \over M}} \; r^2 \sin^2 \ze \; + \; n
(\tau + n \sqrt{M}) \; \left[ \dis{{{r^2 \sin^2 \ze} \over {\sqrt{M} (\tau +
\sqrt{M})}}} - 1 \right] \right\} \ ,
$\hfill} (5b)

\

\hfill{$
(\Hbf_{X \BB_n})_r  \ = \ -\dis{{{\cos \ze} \over {\mu_0 c}}} \;
(\Ebf_{X \BB_n})_\phi \ ,
$\hfill} (5c)

\

\hfill{$
(\Ebf_{X \BB_n})_\phi  \ = \ \dis{{{\cos \ze} \over {\mu_0 c}}} \;
(\Ebf_{X \BB_n})_r \ ,
$\hfill} (5d)

\

\hfill{$
(\Ebf_{X \BB_n})_z  \ = \ \dis{{{\sin^2 \ze} \over {M^2}}} \; \left[ (n^2 - 1) M +
3 \tau (\tau + n \sqrt{M}) \right] \; \Phi_{X \BB_n} \ ,
$\hfill} (5e)

\

with \ $n = 0, 1, 2, \ldots$ \  Since Maxwell equations are linear, both real
and imaginary parts of their solutions are also solutions. In the following,
we shall consider the real part only.\\

{\em B. -- Pointing Flux and Energy Density}

\h From eqs.(5a)-(5e), one obtains[24] the Poynting flux $\Pbold_{X \BB_n}$
and the energy density $U_{X \BB_n} = \eps_0 |\Rerm \Ebf_{X \BB_n}|^2 +
\mu_0 |\Rerm \Hbf_{X \BB_n}|^2$ of the $n\/$th-order localized (limited-diffraction)
electromagnetic X-waves [$n = 0, 1, 2, \ldots$]:

\

\hfill{$
(\Pbold_{X \BB_n})_r \ = \ \Rerm (\Ebf_{X \BB_n})_\phi \; \Rerm (\Hbf_{X \BB_n})_z \ ,
$\hfill} (6a)

\hfill{$
(\Pbold_{X \BB_n})_\phi \ = \ - \Rerm (\Ebf_{X \BB_n})_r \; \Rerm (\Hbf_{X \BB_n})_z \ ,
$\hfill} (6b)

\hfill{$
(\Pbold_{X \BB_n})_z \ = \ {{\cos \ze} \over {\mu_0 c}} \; \left[
|\Rerm (\Ebf_{X \BB_n})_r|^2 + |\Rerm (\Ebf_{X \BB_n})_\phi|^2 \right] \ ,
$\hfill} (6c)

and

\hfill{$
U_{X \BB_n} \ = \ \eps_0 (1 + \cos^2 \ze) \; \left[
|\Rerm (\Ebf_{X \BB_n})_r|^2 + |\Rerm (\Ebf_{X \BB_n})_\phi|^2 \right] +
\mu_0 |\Rerm (\Hbf_{X \BB_n})_z|^2 \ .
$\hfill} (7)

\

The total energy of the $n\/$th-order localized electromagnetic X-waves is
given by

\

\hfill{$
U^{\rm tot}_{X \BB_n} \ = \  \int_{-\pi}^\pi d\phi \ \int_{-\infty}^\infty
dz \ \int_0^\infty r dr U_{X \BB_n} \ ,
$\hfill} (7')

\

which ---at least for the solutions described in this paper--- is infinite
because in the present case the decay of the energy density along the X
branches[4,16] approaches \ $1 / |z-{c \over {\cos \ze}}t|$. \ This is just
the same situation met when using plane waves (which also correspond to
infinite total energies, but finite energy densities). \ And, as in the case
of plane waves, any realistic X-wave (approximately produced {\em with a
finite aperture}) will possess a finite energy[5]. We shall come back to this
question. \ Let us mention that ---however--- Special (Extended) Relativity
[see Sect.V, below] predicts the existence also of finite-energy
(non-truncated) X-shaped waves[39]. \ (Moreover, finite energy, localized
beams can a priori be obtained even via suitable superpositions of X-waves,
i.e., by suitable superpositions of Bessel beams).

\h As already mentioned, the beams represented by eq.(4) can be
particle-like. As the scaling parameter, $a_{0}$, decreases, the
wave-function energy density around the wave centre increases. The
off-center energy density of the X-shaped waves decays slowly along the X
branches, which might provide a way to communicate with other wave
particles[28,40,41].\\

\begin{figure}[!h]
\begin{center}
 \scalebox{0.375}{\includegraphics{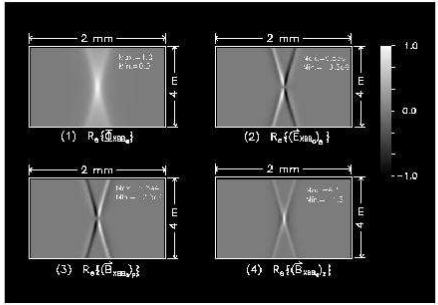}}
\end{center}
\caption{Real part of the Hertz potential and field components of the
zeroth-order ($n=0$) localized electromagnetic X-shaped wave at time
$t=z/c_{1}$. \ Panel (1) is the Hertz potential $\Rerm \{\Phi_{X \BB_0}\}$;
 \ Panel (2) is the $\phi$ component of the electric field strength,
$\Rerm \{(\Ebf_{X \BB_0})_\phi \}$; \ and  Panels (3) and (4) are the $r$
and $z$ components of the magnetic field strength,
$\Rerm \{(\Hbf_{X \BB_0})_r \}$ and $\Rerm \{(\Hbf_{X \BB_0})_z \}$,
respectively. \ The dimension of each panel is 4 m ($r$ direction) ${\times}$
2 mm ($z$ direction). The free parameters $\ze$ and $a_{0}$ are
$0.005^{\rm o}$ and 0.05 mm, respectively. The values shown on the right-top
corner of each panel represent the maxima and the minima of the images
before normalization for display [MKSA units] (see also Table I).}
\label{fig1}
\end{figure}

{\em C. -- An Example}

\h In the following, we give an example of electromagnetic X-shaped wave.
For simplicity, only the zeroth-order ($n=0$) X-wave is considered.
Notice that for $n=0$, eqs.(4')-(5e) are axially symmetric (not a function
of $\phi$), and $(\Ebf_{X \BB_0})_r$, $(\Hbf_{X \BB_0})_\phi$  and
$(\Pbold_{X \BB_0})_\phi$ are zero.

\begin{figure}[!h]
\begin{center}
 \scalebox{0.375}{\includegraphics{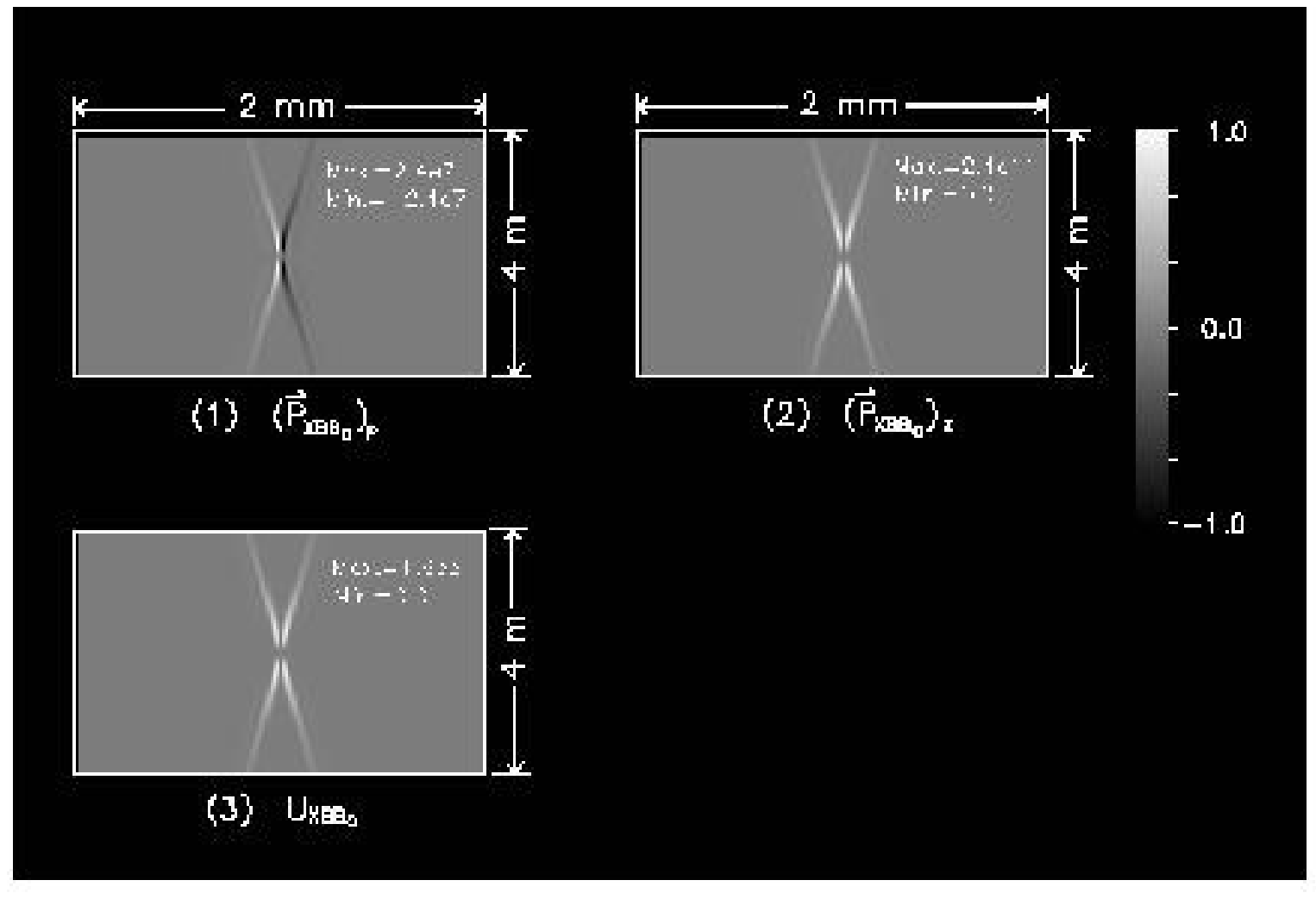}}  
\end{center}
\caption{Poynting flux and energy density of the zeroth-order localized
electromagnetic X-shaped wave at time $t=z/c_{1}$. Panels (1) and (2) are
the $r$ and $z$ components of the Poynting flux, $(\Pbold_{X \BB_0})_r$
and $(\Pbold_{X \BB_0})_z$, respectively; and Panel (3) is the energy
density $U_{X \BB_0}$. \ The dimension of each panel and the parameters of
the X-waves are the same as those in Fig.1. The values shown on the
right-top corner of each panel represent the maxima and the minima of
the images before normalizing for display [MKSA units] (see also Table
II).}
\label{fig2}
\end{figure}

\begin{figure}[!h]
\begin{center}
 \scalebox{0.3}{\includegraphics{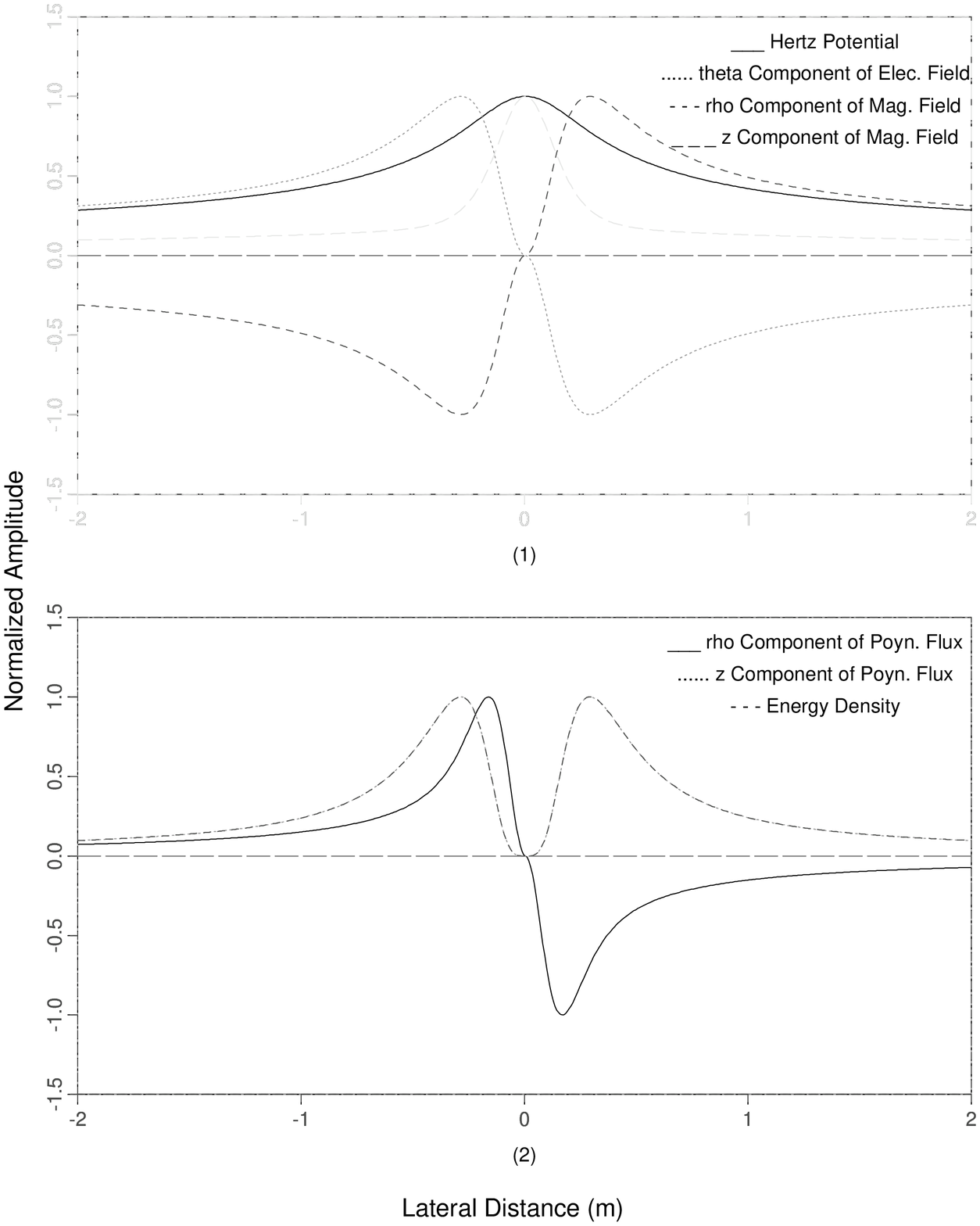}}  
\end{center}
\caption{Line plots of the zeroth-order electromagnetic X-shaped wave
in Figs.1 and 2 along one of the ``X" branches (from left bottom to top
right). \ Panel (1) shows the line plots of the field components: \ $\Rerm
\{\Phi_{X \BB_0}\}$ {\em (full line)\/}; \ $\Rerm \{(\Ebf_{X \BB_0})_\phi
\}$ {\em (dotted line)\/}; \ $\Rerm \{(\Hbf_{X \BB_0})_r \}$ {\em (dashed
line)\/}; \ $\Rerm \{(\Hbf_{X \BB_0})_z \}$ {\em (long-dashed line)\/}. \
Panel (2) is the line plot of the Poynting flux and energy density: \
$(\Pbold_{X \BB_0})_r$ (full line); \ $(\Pbold_{X \BB_0})_z$ {\em (dotted
line)\/}; \ and \ $U_{X \BB_0}$ {\em (dashed line)\/}.}
\label{fig3}
\end{figure}

\h The real part of the zeroth-order scalar Hertz potential
(eq.(4')) and the electromagnetic X-shaped waves (eqs.(5b),(5c) and
(5e)) are shown in Fig.\ref{fig1}.  The Poynting flux and energy density are shown
in Fig.\ref{fig2}.  Their maxima and minima are summarized in Tables I and II,
respectively.  From Table I, we see that, with the parameters given in
Fig.1, the electromagnetic X-shaped waves are almost transverse waves where
their axial components are much smaller than those of the transverse
components.  This is also shown in Table II, where the axial component of
the Poynting flux is at least four orders larger than its lateral
components.  Lateral line plots of Figs.1 and 2 along X branches are
shown in Fig.\ref{fig3}.\\

{\em D. -- Finite-Aperture Approximation of X-Shaped Waves and their
Depth of Field}.

\h The localized electromagnetic X-shaped waves obtained above by us are
exact solutions to the free-space Maxwell wave equations. In these equations,
there are no boundary conditions and thus the apertures required to produce
the waves are infinite; therefore, they cannot be realized in practice.
However, these waves can be approximated very well, over a certain ``depth
of field", by truncating them in both space and time.

\h One important question is how far the truncated X-shaped waves can
travel without appreciable distortion, i.e., how long is their field
depth. The answer is that they travel practically undistorted along a
{\em large} depth of field, and then they suddenly decay.  This has been
shown, even experimentally, in the acoustic case[4-6,16,17], and
mathematically in the more general case of the slingshot pulses
(X-shaped--type waves, found independently in ref.[23] for the case of
homogeneous, general scalar wave-equations) as well as for other
pulses[15,26]. \ Let us address the same question in the present
(electromagnetic) case.

\h Since \ $|\Ebf_{X \BB_n}(r,\phi,z-c_1t)| \ll |\Ebf_{X \BB_n}(r,\phi)|$ \
and \ $|\Hbf_{X \BB_n}(r,\phi,z-c_1t)| \ll |\Hbf_{X \BB_n}(r,\phi)|$ \ for
$|z-c_1t| > d_z/2$ \ within a finite transverse aperture, where $d_z$ is a
constant quantity, the X-shaped waves can be truncated e.g. within the axially
moving window $[c_1t-d_z/2, \ c_1t+d_z/2]$. The truncated waves do not meet
the problems of the theoretical (infinitely extended) ones. \ If the diameter
of the aperture is $D$, the depth of field (which, following Durnin, is
defined as the axial distance at which the field amplitude falls to half of
that at the surface of the source) of the X-wave Hertz potential
(eqs.(4') and (4'')) is given by[4,16,24,42]

\

\hfill{$
Z_{\rm max} \ = \ \dis{{D \over 2} \; \dis{{1 \over {\sqrt{({c_1 / c})^2 - 1}}}}}
 \ = \ \dis{{D \over 2}} \; \cot \ze \ .
$\hfill} (8)

\

Because the derivatives in eqs.(3') and (3'') do not change the cone angle
$\ze$, the field depth of the electromagnetic X-waves is the same as that of
the Hertz potential produced by the same aperture.  In addition, eq.(8) is
also valid for band-limited electromagnetic X-waves[16]. \ As an example,
if the diameter of the aperture is 20 m and the cone angle is $\ze =
0.005^{\rm o}$, the depth of field of both broadband and band-limited
electromagnetic X-waves is 115 km. \ For instance, simulation[4] of a finite
aperture X-wave Hertz potential with the Rayleigh-Sommerfeld diffraction
formula[36] and its production[5] with an acoustic transducer have been
reported in the past literature.

\h Another, even more important question is whether the truncated X-wave
(after having been produced, and without {\em further} support from the
finite antenna) does still travel with Superluminal group-velocity, at least
along its depth of field. \ One may observe that any wave packet of the type
of eq.(4), even with
{\em different} weights (as far as only $z$ depends on $t$), does not spread
in the energy--impulse space; \ in other words, there is not spreading in
the energy--impulse space of the wave packet even after its truncation, if
the linearity between ${\Ecal}$ and $|\imp|$ (i.e., between the coefficients
of $z$ and of $t$ in the exponential phase-factor entering eq.(4)) is
maintained. \ Therefore, after its emission  (even if in a {\em truncated}
form, implying a distortion in the ordinary space, but not in the
energy--impulse space), the wave packet's energetic spectrum does not change,
and thus its group velocity $d{\Ecal} / dk$ is not expected to change too. \
We may conclude that, till when the group-velocity has a meaning (i.e., till
the abrupt decay of the truncated X-wave), $v_\grm$ is expected to remain
Superluminal.\\

{\em E. -- A digression: Localized solutions to the free Schroedinger
equation}.

\h We already mentioned that non-diffractive solutions to the quantum
equations, including Klein-Gordon's and Dirac's, being localized and
particle-like, can well be related to elementary particles (and their wave
nature)[28,43]. \ Let us here make a brief digression with respect to the
particular case of the Schroedinger equation which, being non-relativistic,
presents different problems.

\h It is well-known[44] that the time-independent Schroedinger and Helmholtz
equations are mathematically identical (so that, for instance, the tunnelling
of a particle through and under a barrier has been simulated by the
travelling of photons in a subcritical waveguide). \ Those equations are
however different (due to the different order of the time derivative) in
the time-dependent case. Nevertheless, it can be shown that they still
possess in common classes of analogous solutions, differing only in their
spreading properties[45]. \ We are going to see an example of this fact in the
particularly simple case of free space.

\h The three-dimensional, time-dependent Schroedinger wave equation for
free objects is given by

\

\hfill{$
- \dis{{\hbar \over {2m}}} \; \nabla^2 \Phi \ = \ i \dis{{{\pa \Phi} \over {\pa t}}} \ ,
$\hfill}

\

where $\Phi = \Phi(r,\phi,z,t)$ is now the wave-function. \ It is
easy to verify that eq.(3) is an exact solution also of such
equation, if $f(s) = e^s$, with \ $s = a_0(k,\ze)r
\cos(\phi-\theta) + b(k,\ze)[z-c_1(k,\ze)t]$ \ and \ $c_1(k,\ze) =
\hbar[\al_0^2(k,\ze)+b^2(k,\ze)] / [i2m \, b(k,\ze)]$. \ The
quantities $\al_0(k,\ze)$ and $b(k,\ze)$ are any functions of the
free parameters $k, \ \ze$, while $k \equiv 2\pi/\la$ it the
wave-number and $\la$ the wavelength.

\h Let \ $\al_0(k,\ze) = -ik \, \sin\ze$ \ and \ $b(k,\ze) = ik \,
\cos\ze$; \ we then obtain \ $c_1(k,\ze) \; = \; \hbar k / [2m
\cos\ze] \; = \; v / [2 \cos\ze]$, \ where we used the de Broglie
relation $\hbar k = h / \la = p$, with $p = mv$, the quantity $v$
being the particle speed. \  By using these last relations, and by
putting $f(s) = s$, \ $T(k) = B(k) \, e^{-a_0 k}$, \ and \
$A(\theta) = i^n e^{in\theta}$ into eq.(3), we obtain[24] an
$n\/$th-order localized solution to the free Schroedinger equation
[$n = 0, 1, 2, \ldots$]

\

\hfill{$
\Phi_{X_n} \ = \
 e^{in\phi} \; \int_0^\infty \; B(k) J_n(kr \sin \ze) e^{-k [a_0 - i
 \cos \ze (z - {v \over {2\cos\ze}}t)]} \; dk \ ,
$\hfill} (9)

\

\h However, in the present non-relativistic case, the integration variable $k$
enters the exponential non-linearly (due to the presence of $v$), so that
in eq.(9) {\em the integration cannot be performed in general} (a trivial
exception being, e.g., the case when the weight $BJ_n$ is gaussian). \ The
same non-linear relation between the coefficients of $z$ and $t$ in the
phase factor implies (at variance with the relativistic case) the existence
of a spreading.

\h The group velocity is still given, as usual, by the relation \ $\pa [k z
\cos\ze - k v t /2] \, / \, \pa k \; = \; 0$,\ with \ $v = \hbar k / m$ \ But
now it is necessary to introduce[46] a mean speed, $\ove{v}$, and only at the
price of such an averaging one can write (even in presence of the spreading)
 \ $v_\grm = {\ove{v}}/\cos\ze$. \ Quantity $v_\grm$ can be either slower or
faster than $v$ (and even than $c$).

\h If $\ze = 0, \ a_0 = 0$, and $B(k') = \delta(k'-k)$, eq.(9) represents
just a plane wave propagating in the $z$-direction.

\h For similar work in connection with the Schroedinger equation, see e.g.
Barut, refs.[11,28] and Ignatovich[28]. \ For the relativistic case, see
e.g. refs.[47].\\

\

{\bf IV. -- DISCUSSION}\\

{\em A. -- Hertz vector potentials and $n\/$th-order X-shaped waves}

\h The scalar X-shaped wave $\Phi_{X \BB_n}$ in eq.(4') satisfies the wave
equation (2). It is the component of a Hertz vector potential in the $z$
direction. If $n = 0$, quantity $\Phi_{X \BB_0}$ is axially symmetric and
has a single peak at the wave center. For $n > 0$, it is zero on the
$z$-axis and is axially asymmetric. (When both $n$ and $\ze$ are zero,
$\Phi_{X \BB_n}$ represents the limiting case of a plane wave.) \ The Hertz
potential $\Phi_{X \BB_n}$, which may or may not have a physical meaning,
has been used above as an auxiliary function from which new electromagnetic
X-waves are derived, through eqs.(3') and (3''). \ If $\Phi_{X \BB_n}$ is
treated approximately as one component of $\Ebf$ or of $\Hbf$, it has a
physical meaning: Many microwave and optical phenomena are treated this way
under suitable conditions[36].

\h From our Hertz vector potential, $\Phibf = \Phi {\hat{\zbf}}$, the
family of electromagnetic X-waves, eqs.(5a)-(5e),
has been obtained above, the nonnegative integer $n$
representing the order of the waves. Because the variable $\phi$ appears in
$\Phi_{X \BB_n}$ only, $\Ebf$ and $\Hbf$ have the same axial symmetry as
$\Phi_{X \BB_n}$. \ However, for $n=1$, $\Ebf$ and $\Hbf$ are not zero on
the axis $z$, and they are not axially symmetric.  This means they are not
single-valued on the $z$-axis and thus it does not seem they can be
approximated with a physical device.

\h For some components of the electromagnetic X-waves the field on the
axial axis $z$ is zero. In such a case, there are multiple peaks around the
X-wave center, and the energy density is high on these peaks. For example,
the energy density of the zeroth-order electromagnetic X-wave has four
sharp peaks (see Panel (3) of Fig.2). This is similar to the case of
Ziolkowski's localized electromagnetic wave[21]. Notice that any
higher-order ($n>0$) scalar X-shaped waves are zero on the $z$-axis and
also produce multiple peaks[4].

\h For the electromagnetic X-waves given by eqs.(5a)-(5e), \
$(\Ebf_{X \BB_n})_\phi / (\Ebf_{X \BB_n})_r$ \ or \
$(\Hbf_{X \BB_n})_\phi / (\Hbf_{X \BB_n})_r$ \ is not a function of $r$ and
$\phi$. \ In addition, the $z$-component is very small as compared to the
other two components (see Table I). This means that the major component
of the Poynting flux (eqs.(6a)-(6c)) is in the $z$ direction (the direction
of the vector Hertz potential) (see Table II).

\h One can make recourse to {\em other} Hertz vector potentials, like $\Phibf =
\Phi {\hat{\rbf}}$ \ or \ $\Phibf = \Phi {\hat{\phibf}}$. \ In the latter case,
by using eq.(4'), we obtain electromagnetic X-waves whose
electric components stay in the radial (r,z) plane. \ Similarly, in the
former case, if $\Phi$ is still an X-wave solution, we obtain localized
electromagnetic X-waves with their electric components perpendicular to
$\hat{\rbf}$. \ In all these cases, however, for lower-order $\Phi_{X \BB_n}$
 \ (small values of $n$), \ $\Ebf$ or $\Hbf$ may be singular around the axial
axis because of terms of the type $1/r$ and $1/r^2$.\\

{\em B. --  Method to obtain other limited-diffraction electromagnetic waves}

\h We have seen that, when new localized solutions to eq.(2) are found,
the corresponding localized electromagnetic waves can be obtained [cf. for
instance eqs.(3')-(3'')]. \ There are many ways to obtain new localized
solutions to the scalar wave equation, such as the variable substitution
method, that converts any existing solutions to a localized solution[48];
and the superposition method, that uses Bessel beams or rather X-waves
as basis functions to construct localized beams of practical
usefulness[42].

\h Let us here add the observation that the sidelobes of the scalar X-shaped
waves (eq.(4)) are high along the X branches[16]. The asymptotic behaviour
of the electromagnetic X-wave sidelobes are similar to that of the scalar
waves (see Figs.1 and 2). \ Low sidelobes are necessary for example to
obtain high contrast imaging. To reduce sidelobes of the scalar X-waves in
pulse-echo imaging, recently, unsymmetrical limited diffraction beams such
as bowtie X-waves were developed in acoustics[16]: These waves are obtained
by taking derivatives in one direction, say $y$, of the zeroth-order
X-wave; \  when a bowtie X-wave is used to transmit and its $90^{\rm
o}$-rotated beam pattern (around the $z$-axis) is used to receive,
sidelobes of pulse-echo imaging systems can be reduced dramatically without
compromising the image frame rate. \ The same technique could also be
applied to the electromagnetic X-shaped waves for low sidelobe imaging.\\

{\em C. -- Limited Dispersion and Wave Speed}

\h Because $\Ebf$ or $\Hbf$ in eqs.(3')-(3''), and in similar equations are
obtained by derivatives of the scalar X-waves in terms of their spatial
and time variables, the propagation term $z-c_{1}t$ is retained after the
derivatives. Therefore, the electromagnetic X-shaped waves are also localized
beams: that is, travelling with the wave at speed $c_{1}$, one will see a
constant wave pattern in space.

\h We have seen that the wave speed $c_{1}$ of localized beams (eq.(1))
along the $z$ axis is greater than or equal to the speed of light (in
electromagnetism or optics), or to the speed of sound (in acoustics). For
example, the speed of the X-shaped waves is \ $c_1 = c / \cos\ze \ge c$, \
where $0 \le \ze < \pi/2$ is the cone (Axicon) angle. This behaviour is
said to be ``tachyonic" or Superluminal[39], and has been studied by many
investigators in both ``particle"[7,8,49,50] and wave [4-11,23-28,49]
physics. \ In particular, recently (see e.g. refs.[9,23]) it has been
mathematically shown that all relativistic (homogeneous) wave equations
admit also sub- and Super-luminal solutions: a claim theoretically
confirmed, e.g., in ref.[10] (and that in the past had been verified only
in some particular cases[7]).

\h Although the theoretical Superluminal waves encountered in this
paper cannot be exactly produced, due to their infinite energy, they can
be approximated with a finite aperture radiator and retain ---as we have
seen above--- their essential characteristics (limited-diffraction and
Superluminal group velocity) over a large depth of field. Each component
(wavelet) of the approximated wave propagates at the speed of light (or
of sound), but the conic, or X-shaped wave, created by their superposition
travels at a speed greater than the light (or sound) speed.

\h The X-waves, however, do not appear to violate[7,39,49] the special
theory
of relativity, as we are going to discuss below in Section V, when we shall
also show that Superluminal X-shaped waves, in particular, are predicted by
Relativity itself to travel in space {\em rigidly}, without deforming[7,39]. \

\h Since Superluminal motions seem to appear even in other sectors of
experimental science, we deem it proper to present in an Appendix at the
end of this paper some information about those experimental results. In
fact, the subject of Superluminal objects [or tachyons] addressed in
this paper is still regarded frequently as unconventional, and in this
respect it can get therefore more support from experiments than from theory.
 \ Moreover, such pieces of information are
presently scattered in four different areas of science, and it can be
useful to find them all collected in one and the same place.\\

{\em D. -- Role of superposition and interference}

\h The X-waves considered by us are a superposition of Bessel beams and
therefore can be regarded, eventually, as linear superposition of $c$-speed
plane waves travelling at different angles about the beam axis. If the
aperture is infinite, plane waves will not deform and diverge as they
propagate to infinity. \ If the aperture is finite, plane waves travel to a
large distance in which they do not diverge significantly: However, beyond
that large distance, plane waves, so the X-waves, start to diverge. \ The
speed of the X-waves along the axis, that is greater than $c$, is caused by
the plane waves that do not travel along the axis but at an angle ({\em
Axicon angle\/}) with it.

\h If the aperture is infinite, X-waves travel to infinite
distance without deformation: This is because antennas continue to support
the waves from the X branches which extend to infinity in time as the
aperture goes to infinity (remember that the X-wave extension in time
is equal to $(D/2) \, \tan\ze$, where $D$ is the diameter of the aperture and
$\ze$ is the cone angle).

\h If the X-waves are truncated, they are no longer X-waves in the original
sense.
As soon as truncated X-waves leave the source that produces them, they start
deforming progressively. We have seen that the truncated X-waves deforms
negligibly within their depth of field and significantly beyond it; the peak
of the truncated X-waves happens near the wave axis: it travels faster than
$c$. Beyond the support (depth of field) of the X branches, truncated waves
start to diverge quickly. In the limit (infinite distance), truncated X-waves
behave like a spherical wave travelling at speed $c$. In conclusion,
truncated X-waves can send signals at a speed greater than $c$ only
within the support of the X branches: but this is achieved at the cost that
the wave front must be leaning forward in space to form the advanced
X branches. \ Namely, the outmost ring of the finite antenna has to be
excited first; then, since the waves from this ring travel at $c$, the peak
of the X-waves travels faster than $c$ on the axis. \ The waves deform quite
dramatically near the boundary of the depth of field $d = a \cot \ze$,
where $a$ is the radius of the circular aperture and $\ze$ is the cone
angle.

\h Notice that, however, the peak (vertex) of the truncated X-wave is
formed with wave components from the X branches in a continuously changing
way as the beam propagates, so that it does not represent the speed of the
centroid of the beam. \ This is particularly true when the arms carry large
amounts of energy.\\

\

{\bf V. -- A THEORETICAL FRAMEWORK WITHIN SPECIAL RELATIVITY FOR THE
X-SHAPED WAVES}\\

\h Let us here mention that a simple theoretical framework exists[7]
(merely based on the space-time geometrical methods of Special
Relativity (SR)) which incorporates the Superluminal X-shaped waves
without violating the Relativity principles.

\h Actually, SR can be derived from postulating: (i) the Principle of
relativity; and (ii) space-time to be homogeneous and space isotropic. \
It follows that one and only one {\em invariant}  speed exists; and
experience shows that invariant speed to be the one, $c$, of light in vacuum
(the essential role of $c$ in SR being just due to its invariance, and not
to its being supposedly a maximal, or minimal, speed; no other, sub- or
Super-luminal, object can be endowed with an invariant speed: in other words,
no bradyon or tachyon can play in SR the same essential role as the
speed-$c$ light waves). Let us recall, incidentally, that tachyon
[a term coined in 1967 by G.Feinberg] and bradyon [a term coined in 1970
by E.Recami] mean Superluminal and subluminal object, respectively. The
speed $c$ turns out to be a limiting speed:
but any limit possesses two sides, and can be approached a priori both
from below and from above. \ (As E.C.G.Sudarshan put it, from the fact that
no one could climb over the Himalayas ranges, people of India cannot
conclude that there are no people North of the Himalayas...; actually,
speed-$c$  photons exist, which are
born, live and die just ``at the top of the mountain,"
without any need to accelerate from rest to the light speed).

\h A consequence is that the quadratic form \ $ds^{2} = c^{2} dt^{2} - dx^{2}
- dy^{2} - dz^{2}$ \ (i.e., the four-dimensional length-element square,
along the space-time path of any object) results to be invariant, {\em except
for its sign}. \ In correspondence with the positive (negative) sign, one
gets the subluminal (Superluminal) Lorentz transformations [LT]. The
ordinary, subluminal LTs leave, e.g., the fourvector squares and the
scalar products (between fourvectors) just invariant.

\begin{figure}[!h]
\begin{center}
 \scalebox{0.8}{\includegraphics{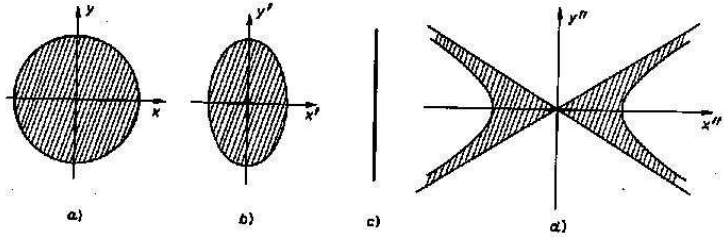}}  
\end{center}
\caption{Let us consider an object that is intrinsically
spherical, i.e., that is a sphere in its rest-frame (Panel (a)).
After a generic subluminal Lorentz transformation (LT) along $x$,
i.e., under a subluminal $x$-boost, it is predicted by special
relativity (SR) to appear as ellipsoidal due to Lorentz
contraction (Panel (b)).  After a Superluminal $x$-boost (namely,
when this object moves with Superluminal speed $V$), it is
predicted by extended relativity (ER) to appear[7,39] as in Panel
(d), i.e., as occupying the cylindrically symmetric region bounded
by a two-sheeted rotation hyperboloid and an indefinite double
cone. The whole structure is predicted by ER to move {\em
rigidly\/}[39] and, of course, with the speed $V$, the cotangent
square of the cone semi-angle being $(V/c)^2 - 1$. \ Panel (c)
refers to the limiting case when the boost-speed tends to $c$,
either from the left or from the right. (For simplicity, a space
axis is skipped).} \label{fig4}
\end{figure}

\h The Superluminal LTs can be easily written down only in two
dimensions (or in six, or in eight, dimensions...). But they must have
the properties of changing sign, e.g., to the fourvector squares and to
the fourvector scalar products. This is enough to deduce ---see
Fig.\ref{fig4}--- that a particle, which is spherical when at rest (and which
appears then as ellipsoidal, due to Lorentz contraction, at subluminal
speeds $v$), will appear[7,39,40,51]
as occupying the cylindrically symmetrical region bounded by a
two-sheeted rotation hyperboloid and an indefinite double cone, as in
Fig.4(d), for Superluminal speeds $V$. In Fig.4 the motion is along the
$x$-axis. In the limiting case of a point-like particle, one obtains only a
double cone. In 1980-1982, therefore, it was predicted[39] that the simplest
Superluminal object appears (not as a particle, but as a field or rather) as
a wave: namely, as a ``X-shaped wave", the cone semi-angle $\al$ being given
(with $c=1$) by \ $\cot \al = \sqrt{V^2 - 1}$.

\begin{figure}[!h]
\begin{center}
 \scalebox{0.5}{\includegraphics{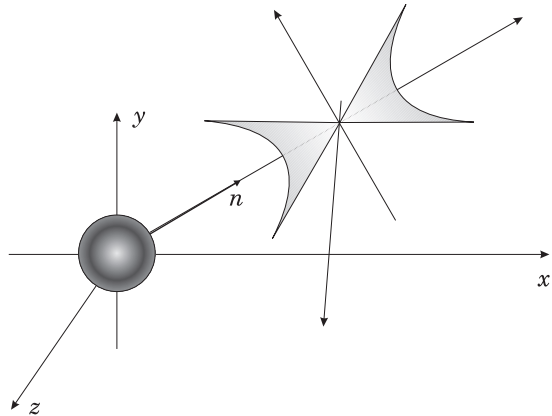}}  
\end{center}
\caption{If we start from a spherical particle as in Fig.4(a), then
---after a Superluminal boost along a generic motion line $l$--- we obtain
the tachyonic object $T$ depicted in this figure. \ Once more, the
Superluminal object $T$ appears to be spread over the whole spatial region
delimited by a double cone and a two-sheeted hyperboloid asymptotic to the
cone[51]. The whole structure travels of course along $l$ with the speed
$V$ of the Superluminal $l$-boost. \ Notice that, if the object
is not spherical when at rest (but, e.g., is ellipsoidal in its own
rest-frame), then the axis of $T$  will no longer coincide with $l$, but
its direction will depend on the speed $V$ of the tachyon itself. \ For the
case in which the space extension of the Superluminal object $T$ is finite,
see ref.[40].}
\label{fig5}
\end{figure}

\h It was also predicted[39,51] that the X-shaped waves would move
{\em rigidly}  with speed $V$ along their motion direction
(Fig.\ref{fig5}). The reason for such ``X-waves" to travel without deformation
is quite simple: every X-wave can be regarded at each instant of time as
the (Superluminal) Lorentz transform of a spherical object, which of
course as time elapses moves in vacuum without any deformation.

\h The X-shaped waves here considered are the most simple ones only.
If we started not from an intrinsically spherical or point-like object,
but from a non-spherically symmetric particle, or from a pulsating
(contracting and dilating) sphere, or from a particle oscillating back
and forth along the motion direction, their Superluminal Lorentz
transforms would result to be more and more complicated. The above-seen
X-waves, however, are typical for a Superluminal object, so as the
spherical or point-like shape is typical for a subluminal
particle.

\begin{figure}[!h]
\begin{center}
 \scalebox{0.3}{\includegraphics{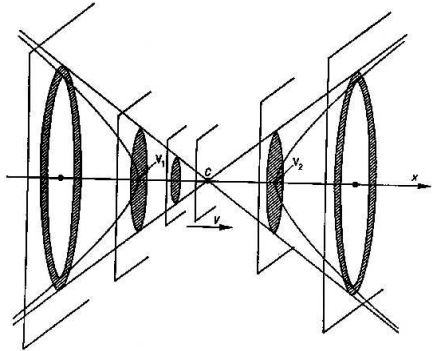}}  
\end{center}
\caption{Here we show the intersections of the Superluminal object
$T$ with planes $P$ orthogonal to its motion line (the $x$-axis) for the
same case as in Fig.4. \ For simplicity, we assumed again the object to be
spherical in its rest-frame, and the cone vertex to coincide with the
origin O for $t=0$. \ Such intersections evolve in time so that the same
pattern appears on a second plane ---shifted by $\Delta x$ after the time
$\Delta t = \Delta x / V$. \ On each plane, as time elapses, the
intersection is therefore predicted by ER to be a circular ring which, for
negative times, goes on shrinking until it reduces to a circle and then to
a point (for $t=0$); \ afterwards, such a point becomes again a circle and
then a circular ring that goes on broadening[7,39,40,51.}
\label{fig6}
\end{figure}

\h The three dimensional picture of Fig.5, or rather of Fig.4(d),
appears in Fig.\ref{fig6}, where its annular intersections with a transverse
plane are shown (cf. refs.[39,51]).

\h It has been believed for a long time that Superluminal objects
would have allowed sending information into the past; but such problems
with causality seem to be solvable within SR. Once SR is generalized in
order to include tachyons, no signal travelling backwards in time is
apparently left.  For a solution of those causal paradoxes, see
refs.[49,50] and references therein.

\h Let us pass, within this elementary context, to the problem of
producing a ``X-shaped wave" like the one depicted in Fig.6
(truncated of course, in space and in time, by use of a finite antenna
radiating for a finite time interval). To convince ourselves about the
possibility of realizing them, it is enough to consider {\em na\"{\i}vely}
the ideal case of a Superluminal source {\em S} of negligible size, endowed
with constant speed $V$ and emitting spherical electromagnetic waves $W$
(each one travelling at the invariant speed $c$).  We shall observe
the electromagnetic waves to be internally tangent to an enveloping cone
{\em C} having the source motion line $x$ as its axis and $S$ as its vertex.
This is analogous to what happens with an airplane moving with a constant
supersonic speed in the air. \ In addition, those
electromagnetic waves {\em W} interfere negatively one another inside the
cone {\em C}, and interfere constructively only on its surface.  We can put
a plane detector orthogonal to $x$ and record the intensity and direction
of the waves {\em W} impinging on it, as a (cylindrically symmetric) function
of position and of time.  Afterwards, it will be enough to replace the plane
detector by a plane antenna that radiates ---instead of detecting and
recording--- exactly the same (cylindrically symmetrical) space-time pattern
of electromagnetic waves {\em W}, in order to build up a cone-shaped
({\em C\/}) electromagnetic wave travelling along $x$ with the
Superluminal speed $V$ (obviously, with no radiating source any longer
---now--- at its vertex {\em S\/}). Even if each spherical wave {\em W} will
still travel with the invariant speed $c$.

\h Incidentally, the above-mentioned evaluation can lead to a very simple
type of ``X-shaped wave."

\h These examples are to be
completed, however, by recalling that Special Relativity implies considering
also the forward cone: cf. Fig.6. \ The truncated X-waves considered in
this paper, for instance, must have a leading cone in addition to the rear cone; such a
leading cone having an essential role for the peak travelling faster than
$c$. \ In other words, in the approximated case in which we
produce a finite conic wave truncated both in space and in time, the
theory of SR suggested the bi-conic shape (symmetrical in space with
respect to the vertex  {\em S}) to be a
priori a better approximation to a rigidly travelling wave; so that SR
suggests to have recourse to an antenna emitting a radiation (not only
cylindrically symmetrical in space but also) symmetric in time, for
a better approximation to an undistorted progressive wave.

\h Let us recall once more that our truncated (finite) bi-conic, or X-shaped,
waves (after having been produced) are expected to travel almost rigidly,
at Superluminal speed, all along their depth of field (but only along
such depth of field).

\h One may observe, at last, that in the vacuum and in homogeneous media for
the (non truncated) X-shaped waves the group velocity coincides with the
phase velocity.\\

\

{\bf VI. -- CONCLUSION}\\

\h We have derived and considered various localized X-shaped solutions
to the free Maxwell equations, which possess Superluminal group velocities.
Theoretically, these solutions are diffraction-free. They are
infinitely extended in space and time, and ---the ones met in this work---
possess infinite total energy. However, for solutions that are single
valued and non-singular (finite energy density), they can be approximated
very well by a finite aperture antenna over a large depth of interest
(along which the truncated X-shaped waves maintain their essential
characteristics of limited-diffraction and Superluminal velocity). \ In
addition, localized, limited-diffraction wave solutions to the quantum
equations can be helpful for a better understanding of the relationship
between waves and particles. \ Because (practically realizable) localized
beams have a large depth of field, they have and can have applications in
several areas, from electromagnetism, communications and optics to acoustics
and geophysics.

\h As explained in Sect.IV--C above, we use the present occasion to
present in the Appendix below some information about the other sectors
of experimental science in which Superluminal motions seem to
appear.\\

\

{\bf VII. -- ACKNOWLEDGMENTS}\\

{\bf We gratefully acknowledge that this paper is based on work developed
in collaboration with Professors Dr. Jian-yu Lu and
Dr. J.F.Greenleaf and appeared in preliminary form as Report INFN/FM--96/01
(I.N.F.N.; Frascati, 1996) \ [its number in the electronic LANL Archives being
\# physics/9610012; \ Oct.15, 1996]; in particular, the Tables and Figs.1--3
are due to Lu and Greenleaf.} \  The mistakes eventually appearing in
this article, however, are the responsibility of the present author only. \
The first three figures of that Report, and of this paper, will be used
---by permission of Lu, Greenleaf and Recami--- also in a paper by
W.A.Rodrigues Jr. and J.-y.Lu (submitted to {\em Found. Phys.\/}). \
The author is very grateful to the co-editor, Prof.D.Stauffer, for many
helpful critical comments. \ He appreciates
for useful discussions, moreover, Dr. Vladislav S. Olkhovsky,
of the Institute for Nuclear Physics of the Ukrainian Academy of Sciences,
Kiev; Dr. Amr M. Shaarawi, of the Faculty of Engineering, Cairo University,
Giza, Egypt; Dr. Peeter Saari, of the Tartu University, Estonia;  and Dr.
Hugo E. Hern\'{a}ndez and Dr. L. C. Kretly, of the
D.M.O.-FEEC and of the C.C.S., respectively, at the Universidade Estadual de
Campinas, Campinas, S.P., Brazil.\\

\

{\bf VIII. -- APPENDIX}\\

\h In this Appendix we take the opportunity to present some sketchy
information ---mainly bibliographical--- about the other three
(in total, four) sectors of experimental science in which Superluminal
motions seem to appear. In fact, as the ``Superluminal" topic
is still controversial, a panoramic view of the overall
{\em experimental}  situation is certainly
useful, especially when it is considered that the related information,
scattered in very different Journals, is not all of easy access to
everybody.

\h The question of Superluminal objects or waves [tachyons] has a long
story, starting perhaps with Lucretius'  {\em De Rerum
Natura.}   Still in pre-relativistic times, let us recall
e.g. the contributions by A.Sommerfeld.  In relativistic times, our
problem started to be tackled again essentially in the fifties and
sixties, in particular after the papers by E.C.George Sudarshan et al.,
and later on by E.Recami, R.Mignani et al. [who by their numerous works
at the beginning of the seventies rendered, by the way, the terms sub-
and Super-luminal of popular use], as well as by H.C.Corben and others
(to confine ourselves to the theoretical researches).  For references,
one can check pages 162-178 in ref.[7], where about 600 citations are
listed; or the large bibliographies by V.F.Perepelitsa[52] and the book
in ref.[53].  In particular, for the causality problems one can see
refs.[49,50] and references therein, while for a model theory for
tachyons in two dimensions one can be addressed to refs.[7,8].  The
first experiments looking for tachyons were performed by T.Alvager et
al.; some citations about the early experimental quest for Superluminal
objects being found e.g. in refs.[7,54].

\h The subject of tachyons is presently returning after fashion,
especially because of the fact that four different experimental sectors
of physics seem to indicate the existence of Superluminal objects;
including ---of course--- the one dealt with in this and in
other papers of ours: which appears to us as being  {\em at the
moment}  the most important sector.  Let us put forth in the
following some brief information about the experimental results obtained
in such different science areas.\\

{\em  1. -- Negative square-mass neutrinos}

Since 1971 it was known that the experimental square-mass of
muon-neutrinos resulted to be negative[55]. If confirmed, this would
correspond (within the ordinary naive approach to relativistic
particles) to an imaginary mass and therefore to a Superluminal speed.
[In a non-naive approach[7], i.e. within a Special Relativity theory
extended to include tachyons (Extended Relativity), the free tachyon
``diffraction relation" (with $c=1$) becomes actually \ $E^2 - \imp^2 =
- m_0^2$, so that one does not have to associate tachyons with
imaginary-mass particles!].

\h From the theoretical point of view, let us refer to [56] and
references therein.

\h It is rather important that recent experiments showed that also
electron-neutrinos have negative mass-square[57].\\

{\em 2. -- Galactic ``Mini-Quasars"}

We refer ourselves to the apparent Superluminal expansions observed
inside quasars, some galaxies, and ---as discovered very
recently--- in some galactic objects, preliminarily called
``mini-quasars". Since 1971 in many quasars
apparent Superluminal expansions were observed
[{\em Nature}, for instance, dedicated a cover of its to
those observations on 2 Apr.1981]. Such seemingly Superluminal
expansions were the consequence of the experimentally measured angular
separation rates, once it was taken account of the (large) distance of
the sources from the Earth. From the experimental point of view, it will
be enough to quote the book [58] and references therein.

\h The distance of those ``Superluminal sources", however,
is not well-known; or, at least, the (large) distances usually
adopted have been strongly criticized by H.Arp et al., who maintain that
quasars are much nearer objects than expected: so that all the
above-mentioned data can no longer be easily used to infer (apparent)
Superluminal motions.  However, very recently,
 {\em galactic}  objects have been discovered,
in which apparent Superluminal expansions occur; and the distance of the
galactic objects can be more precisely determined. For the experimental
data, see refs.[59].

\h For a theoretical point of view, both for quasars and
``mini-quasars", see refs.[7,60]. In particular, let us recall
that a {\em single}  Superluminal source of
light would just be observed:  (i) initially, in the phase of ``optic
boom" (analogous to the acoustic ``boom" by an aircraft
that travels with constant super-sonic speed) as a suddenly-appearing
intense source;  (ii) later on, as a source which splits into
TWO objects receding one from the other with relative velocity
{\em V} larger than {\em 2c.}\\

{\em 3. -- Tunnelling photons = Evanescent waves}

It is the sector that most attracted the attention of the
scientific and non-scientific {\em press\/}[61].

\h Evanescent waves were predicted [cf., e.g., ref.[7], page 158 and
references therein] to be faster-than-light. Even more, they consist in
tunnelling photons: and it was known since long time[62] that tunnelling
particles (wave packets) can move with Superluminal group--velocities
inside the barrier; therefore, due to the theoretical analogies between
tunnelling particles (e.g., electrons) and tunnelling photons[44], it
was since long expected that evanescent waves could be Superluminal.

\h The first experiments have been performed in Cologne, Germany, by
Guenter Nimtz et al., and published in 1992. See refs.[63].

\h Important experiments have been performed in Berkeley: see
refs.[64].

\h Further experiments on Superluminal evanescent waves have been done
in Florence[65]; while a last experiment (as far as we know) took place
in Vienna[66].

\h From the theoretical point of view, see refs.[62] and references therein;
and refs.[67].\\

{\em 4. -- Superluminal motions in Electrical and Acoustical Engineering ---
The ``X-shaped waves"}

This fourth sector, which this paper is contributing to, seems to
be at the moment (together with the third one) the most promising.

\h Starting with the pioneering work by H.Bateman, it became slowly
known that all the relativistic homogeneous wave equations ---in a
general sense: scalar, electromagnetic and spinorial--- admit solutions
with subluminal group velocities[1]. More recently, also Superluminal
solutions have been constructed for those homogeneous wave equations, in
refs.[2,10] and quite independently in refs.[6,9,23]: in some cases just
by applying a Superluminal Lorentz ``transformation"[7,8].
It has been shown that an analogous situation is met even for
acoustic waves, with the existence in this case of
``sub-sonic" and ``Super-sonic" solutions[3,4]; \
so that one can expect they to exist, e.g., also for seismic wave
equations. (More intriguingly, one might expect the same to be true in the
case of gravitational waves too).

\h Let us recall that the rigidly travelling Supersonic and
Superluminal solutions found in refs.[4,5,23,27] and in this paper
---some of them already experimentally realized--- appear to be
(generally speaking) X-shaped, just as predicted in 1980/1982 by Barut,
Recami and Maccarrone[39].

\h On this regard, from the theoretical point of view, let us quote
pages 116-117, and pages 59 (fig.19) and 141 (fig.42), of ref.[7]; and
even more refs.[7,40,51], where ``X-type waves" are
predicted and discussed.  From such papers it results also clear why the
X-shaped waves keep their form while travelling (non-diffractive
waves).\\

\

{\em Note added:\/} {\bf during the about two years elapsed between
original submission and publication, many relevant papers appeared in print.
Quotations of various of them have been directly included in the References.
After the revision of the citations, however, some further important articles
appeared. Let us mention in particular that: \ (i) at Tartu (Estonia) they
have experimentally produced X-shaped
(Superluminal) light waves, in optics: see P.Saari and K.Reivelt: ``Evidence
of X-shaped propagation-invariant localized light waves", appeared in
{\em Phys. Rev. Lett.}, Nov.24, 1997; \ H.S\~{o}najalg, M.R\"{a}tsep and
P. Saari: {\em Opt. Lett.} 22 (1997) 310; and P.Saari and H.S\~{o}najalg:
{\em Laser Phys.} 7 (1997) 32}; \ (ii) simultaneously, (non-truncated)
X-shaped beams with {\em finite} total {\em energy} ---expected to exist on
the basis of ER--- were mathematically constructed by I.Besieris,
M.Abdel-Rahman, A.Shaarawi and A.Chatzipetros in the work ``Two fundamental
representations of localized pulse solutions to the scalar wave equation",
to appear in {\em J. Electromagnetic Waves Appl.} (1998).\hfill\break

%
%
%

\vfill
\newpage

\centerline{{\bf References}}

\

[1] H.Bateman: {\em Electrical and Optical Wave
Motion}  (Cambridge Univ.Press; Cambridge, 1915); \
A.O.Barut and A.Grant:  {\em Found. Phys. Lett.}
3 (1990) 303;  A.O.Barut and A.J.Bracken: {\em Found. Phys.}  22 (1992)
1267. \ See also refs.[14,19,20] below.\hfill\break

[2] A.O.Barut and H.C.Chandola:  {\em Phys. Lett.}
A180 (1993) 5. See also A.O.Barut: {\em Phys. Lett.} A189 (1994) 277-280,
and A.O.Barut et al.: refs.[1].\hfill\break

[3] Jian-yu Lu and J.F.Greenleaf: {\em IEEE
Transactions on Ultrasonics, Ferroelectrics, and Frequency
Control}  37 (1990) 438-447.\hfill\break

[4] Jian-yu Lu and J.F.Greenleaf: {\em IEEE Transactions
on Ultrasonics, Ferroelectrics, and Frequency Control}
39 (1992) 19-31 (awarded paper). \ Cf. also ref.[23], where X-shaped--type
waves were found and called  ``slingshot pulses".\hfill\break

[5] Jian-yu Lu and J.F.Greenleaf: {\em IEEE Transactions on
Ultrasonics, Ferroelectrics, and Frequency Control}  39
(1992) 441-446 (awarded paper).\hfill\break

[6] R.Donnelly, D.Power, G.Templeman and A.Whalen:
{\em IEEE Transactions on Ultrasonics, Ferroelectrics and
Frequency Control} 41 (1994) 7-12.\hfill\break

[7] E.Recami: ``Classical tachyons and possible
applications,"  {\em Rivista Nuovo
Cimento}  9 (1986), issue no.6, pp.1-178; and
refs. therein.\hfill\break

[8] E.Recami and W.A.Rodrigues: ``A model theory for tachyons in
two dimensions", in  {\em Gravitational Radiation and
Relativity,}  J.Weber and T.M.Karade editors (World
Scient.; Singapore, 1985), pp.151-203.\hfill\break

[9]  See e.g. R.Donnelly and R.W.Ziolkowski: {\em Proc. Royal Soc.
London}  A440 (1993) 541. \ Cf. also I.M.Besieris, A.M.Shaarawi and
R.W.Ziolkowski: {\em J. Math. Phys.} 30 (1989) 1254-1269.\hfill\break

[10] S.Esposito: {\em Phys. Lett}. A225 (1997) 203; \
W.A.Rodrigues Jr. and J.Vaz Jr., ``Subluminal and
superluminal solutions in vacuum of the Maxwell equations and the
massless Dirac equation," to appear in  {\em Adv. Appl.
Cliff. Alg.}\hfill\break

[11] R.W.Ziolkowski: {\em Phys. Rev.} A39 (1989) 2005-2033; \ A44 (1991)
3960-3984; \ R.W.Ziolkowski, D.K.Lewis and B.D.Cook:
{\em Phys. Rev. Lett.} 62 (1989) 147-150; \
R.W.Ziolkowski, I.M.Besieris and A.M.Shaarawi: {\em Proceed. IEEE} 79 (1991)
1371-1378; \ A.M.Shaarawi, I.M.Besieris and R.W.Ziolkowski: {\em Optics
Comm.} 116 (1995) 183-192; \ G.Indebetouw: {\em J. Opt. Soc. Am.} 6 (1989)
150-152; \ P.L.Overfelt: {\em Phys. Rev.} A44 (1991)
3941; \ A.O.Barut: {\em Phys. Lett.} A143 (1990) 349; \
A.O.Barut and A.Grant: in ref.[1]; \ A.O.Barut:
in {\em L. de Broglie, Heisenberg's Uncertainties and the Probabilistic
Interpretation of Wave Mechanics} (Kluwer; Dordrecht, 1990); \ A.O.Barut:
``Quantum theory of single events: Localized de Broglie--wavelets,
Schroedinger waves and classical trajectories", preprint IC/90/99 (ICTP;
Trieste, 1990).\hfill\break

[12] R.Courant and D.Hilbert: {\em Methods of Mathematical Physics}
(J.Wiley; New York, 1966), vol.2, p.760.\hfill\break

[13] J.A.Stratton:  {\em Electromagnetic
Theory}  (McGraw-Hill; New York, 1941), p.356.\hfill\break

[14] J.Durnin: {\em J. Opt. Soc. Am.}  4 (1987) 651-654; \ J.Durnin,
J.J.Miceli Jr. and J.H.Eberly: {\em Phys. Rev. Lett.}  58 (1987)
1499-1501; \ {\em Opt. Lett.} 13 (1988) 79. \ See also P.Saari and
H.S\~{o}najalg: quoted in the {\em Note added in proofs}.\hfill\break

[15] See e.g. A.M.Vengsarkar, I.M.Besieris, A.M.Shaarawi and R.W.Ziolkowski:
{\em J. Opt. Soc. Am. A\/}9 (1992) 937-949; \ S.M.Sedky, A.M.Shaarawi,
I.M.Besieris and F.M.M.Taiel: {\em J. Opt. Soc. Am. A\/}13 (1996) 1719-1727;
 \ R.W.Ziolkowski and D.K.Lewis: {\em J. Appl. Phys.} 68 (1990) 6083; \
R.W.Ziolkowski, I.M.Besieris and A.M.Shaarawi: in ref.[11]; \
A.M.Shaarawi, I.M.Besieris and R.W.Ziolkowski: {\em J. Appl. Phys.} 65 (1989)
805-813. \ See also B.Salik, J.Rosen and A.Yariv: {\em Opt. Lett.} 20 (1995)
1743.\hfill\break

[16] Jian-yu Lu: {\em IEEE Transactions on Ultrasonics, Ferroelectrics and
Frequency Control}  42 (1995) 1050-1063; \ 40 (1993) 735-746.\hfill\break

[17]  Jian-yu Lu, He-hong Zou and J.F.Greenleaf: {\em Ultrasound in
Medicine and Biology} 20 (1994) 403-428.\hfill\break

[18] Jian-yu Lu and J.F.Greenleaf: {\em Ultrasonic Imaging}  15
(1993) 134-149.\hfill\break

[19] J.N.Brittingham: {\em J. Appl. Phys.}  54 (1983) 1179-1189.\hfill\break

[20] R.W.Ziolkowski: {\em J. Math. Phys.}  26 (1985) 861-863. \ Cf.
also R.W.Ziolkowski, D.K.Lewis and B.D.Cook: in ref.[11].\hfill\break

[21] R.W.Ziolkowski: refs.[11].\hfill\break

[22] K.Uehara and H.Kikuchi: {\em Appl. Physics}  B48 (1989) 125-129; \
J.Rosen, B.Salik, A.Yariv and H.-K. Liu: {\em Opt. Lett.} 20 (1995) 423; \
D.Eliyahu, R.A.Salvatore, J.Rosen, A.Yariv and J.-J.Drolet: {\em Opt. Lett.}
20 (1995) 1412.\hfill\break

[23] R.W.Ziolkowski, I.M.Besieris and A.M.Shaarawi: {\em J. Opt. Soc. Am.}
A10 (1993) 75. \ Cf. also R.W.Ziolkowski: in ref.[11].\hfill\break

[24] J.-y.Lu, J.F.Greenleaf and E.Recami: ``Limited-diffraction solutions
to Maxwell (and Schroedinger) equations", Report INFN/FM--96/01 (INFN;
Frascati, 23 Oct.1996); electronic LANL Archives \# physics/9610012; \
Oct.15, 1996.\hfill\break

[25] Hugo Hern\'{a}ndez F., \ A.Pablo L.Barbero and \ E.Recami: (unpublished).
\hfill\break

[26] A.M.Vengsarkar, I.M.Besieris, A.M.Shaarawi and
R.W.Ziolkowski: in ref.[15]. \ See also
I.M.Besieris, A.M.Shaarawi and R.W.Ziolkowski: in ref.[9]; \
A.M.Shaarawi, I.M.Besieris and R.W.Ziolkowski: {\em J.
Math. Phys.} 31 (1990) 2511-2519; \ J.Fagerholm, A.T.Friberg, J.Huttunen,
D.P.Morgan and M.M.Salomaa: {\em Phys. Rev.} E54 (1996) 4347-4352.\hfill\break

[27] H.S\~{o}najalg and P.Saari: {\em Opt. Lett.} 21 (1996) 1162-1164;
 \ P.Saari: in {\em Ultrafast Processes in Spectroscopy}, ed. by A.Svelto
et al. (Plenum; New York, 1996), pp.151-156. \ Cf. also J.Turunen, A.Vasara
and A.T.Friberg: {J. Opt. Soc. Am.} A8 (1991) 282; \ K.Reivelt: Thesis
(Univ. of Tartu; 1995).\hfill\break

[28] A.M.Shaarawi, I.M.Besieris and R.W.Ziolkowski: in ref.[26], especially
Sect.VI; \ {\em Nucl Phys. (Proc.Suppl.)} B6 (1989) 255-258; \ {\em Phys.
Lett.} A188 (1994) 218-224; \ A.O.Barut: {\em Phys. Lett.} A171 (1992) 1-2; \
{\em Ann. Foundation L. de Broglie}, Jan.1994; \ and the last one of
refs.[11]. \ See also V.K.Ignatovich: {\em Found. Phys.} 8 (1978) 565-571; \
A.Garuccio and V.A.Rapisarda: {\em Nuovo Cimento} A65 (1981) 269-297; \
D.Bambusi et al.: ``On the relevance of classical electrodynamics for the
foundations of Physics", Milano Univ. preprint (Dip.Mat,; Milan,
1997?).\hfill\break

[29] Jian-yu Lu: ``Producing bowtie limited diffraction beams with
synthetic array experiment," {\em IEEE Transactions on
Ultrasonics, Ferroelectrics, and Frequency Control}  (in
press), and refs. therein; \ ``Limited diffraction array beams,"
 {\em International Journal of Imaging System and
Technology}  (in press), and refs. therein.\hfill\break

[30] D.K.Hsu, F.J.Margetan and D.O.Thompson: {\em Appl. Phys. Lett.}  55
(1989) 2066-2068; \ J.A.Campbell and S.Soloway: {\em J.
Acoust. Soc. Am.}  88 (1990) 2467-2477.\hfill\break

[31] Jian-yu Lu and J.F.Greenleaf: in  {\em IEEE
Ultrasonics Symposium Proceedings} (91CH3079-1),
vol.2 (1991) pp.1155-1159.\hfill\break

[32] Tai K.Song, Jian-yu Lu and J.F.Greenleaf: {\em Ultrasonic
Imaging} 15 (1993) 36-47.\hfill\break

[33] Jian-yu Lu and J.F.Greenleaf: {\em Ultrasound in Medicine and Biology}
17 (1991) 265-281.\hfill\break

[34] G.S.Kino: {\em Acoustic Waves: Device, Imaging, and
Analog Signal Processing}  (Prentice-Hall; Englewood
Cliffs, NJ, 1987), chapts.2 and 3.\hfill\break

[35] E.Heyman, B.Z.Steinberg and L.B.Felsen: {\em J. Opt. Soc. Am.}
A4 (1987) 2081-2091.\hfill\break

[36] J.W.Goodman: {\em Introduction to Fourier Optics} (McGraw-Hill;
New York, 1968), chapts.2-4. \ See also G. Toraldo di Francia: {\em Onde
Elettromagnetiche} (Zanichelli; Bologna, 1988).\hfill\break

[37] J.H.McLeod: {\em J. Opt. Soc. Am.} 44 (1954) 592.\hfill\break

[38] E.P.Wigner: {\em Phys. Rev.} 98 (1955) 145.\hfill\break

[39] A.O.Barut, G. D.Maccarrone and E.Recami, ``On the shape of
tachyons," {\em Nuovo Cimento} A71 (1982) 509-533. \ See also E.Recami
and G.D.Maccarrone: {\em Lett. Nuovo Cim.} 28 (1980) 151-157; \
P.Caldirola, G.D.Maccarrone and E.Recami: {\em Lett. Nuovo Cim.} 29 (1980)
241-250.\hfill\break

[40] E.Recami and G.D.Maccarrone: {\em Lett. Nuovo Cimento} 37
(1983) 345-352.\hfill\break

[41] Cf., e.g., V.A.Rapisarda: {\em Lett. Nuovo Cim.} 33 (1982) 437-444; \
F.Falciglia, A.Garuccio and L.Pappalardo: {\em Lett. Nuovo Cim.} 34 (1982) 1,
and refs. therein.\hfill\break

[42] Jian-yu Lu: in {\em IEEE 1995 Ultrasonic Symp. Proceed.} (95CH35844),
vol.2 (1995) pp.1393-1397; \ ``Designing limited diffraction beams", in
 {\em IEEE Transactions on Ultrasonics, Ferroelectrics and
Frequency Control}  (in press).\hfill\break

[43] Cf. e.g. A.O.Barut: {\em Phys. Lett.} A143 (1990) 349;
\ {\em Found. Phys.} 20 (1990) 1233; \ P.Hillion:  {\em Phys. Lett.}
A172 (1992) 1; \ A.M.Shaarawi, I.M.Besieris and R.W.Ziolkowski: in ref.[26]; \
in ref.[28]; \ and \ {\em Nuclear Physics} (Proc.Suppl.) B6 (1989) 255.\hfill\break

[44] See, e.g., Th.Martin and R.Landauer: {\em Phys. Rev. A\/}{\bf 45}
(1992) 2611; \ R.Y.Chiao, P.G.Kwiat and A.M.Steinberg: {\em Physica
\/}B{\bf 175} (1991) 257; \ A.Ranfagni, D.Mugnai, P.Fabeni and
G.P.Pazzi: {\em Appl. Phys. Lett.} {\bf 58} (1991) 774. \ See also
A.M.Steinberg: {\em Phys. Rev.} A52 (1995) 32.\hfill\break

[45] A.Agresti, V.S.Olkhovsky and E.Recami: (to be submitted for pub.)
\hfill\break

[46] M.L.Goldberger and K.M.Watson: {\em Collision Theory} (Wiley; New York,
1963).\hfill\break

[47] A.M.Shaarawi, I.M.Besieris and R.W.Ziolkowski: in ref.[26], and refs.
therein; \ and \ in ref.[28].\hfill\break

[48] Jian-yu Lu, He-hong Zou and J.F.Greenleaf: {\em IEEE Transactions on
Ultrasonics, Ferroelectrics and Frequency Control} 42 (1995) 850-853. \
See also A.T.Friberg, J.Fagerholm and M.M.Salomaa: {\em Opt. Comm.} 136
(1997) 207-212.\hfill\break

[49] E.Recami: ``Tachyon kinematics and causality",
{\em Foundation of Physics} 17 (1987) 239-296.\hfill\break

[50] E.Recami: ``The Tolman `Anti-telephone' paradox: Its solution
by tachyon mechanics,"  {\em Lett. Nuovo
Cimento} 44 (1985) 587-593.\hfill\break

[51] G.D.Maccarrone, M.Pavsic and E.Recami: {\em Nuovo Cimento}
B73 (1983) 91-111.\hfill\break

[52] V.F.Perepelitsa: Reports ITEF-100 and ITEF-165 (Institute of
Theoretical and Experimental Physics; Moscow, 1980).\hfill\break

[53] E.Recami (editor):  {\em Tachyons, Monopoles, and
Related Topics}  (North-Holland; Amsterdam,
1978).\hfill\break

[54] See e.g. D.F.Bartlett et al.:  {\em Phys.
Rev.}  D18 (1978) 2253; P.N.Bhat et al.:
 {\em J. Phys.}  G5 (1979) L13. \ Cf. also
A.S.Goldhaber and F.Smith:  {\em Rep. Progr.
Phys.}  38 (1975) 757; L.W.Jones:  {\em Rev.
Mod. Phys.}  49 (1977) 717.\hfill\break

[55] See e.g. E.V.Shrum and K.O.H.Ziock:  {\em Phys.
Lett.}  B37 (1971) 114; D.C.Lu et al.:
 {\em Phys. Rev. Lett.}  45 (1980) 1066;
G.Backenstoss et al.:  {\em Phys. Lett.}  B43
(1973) 539; H.B.Anderhub et al.: {\em Phys.
Lett.}  B114 (1982) 76; R.Abela et al.:
 {\em Phys. Lett.}  B146 (1984) 431;
B.Jeckelmann et al.:  {\em Phys. Rev. Lett.}
56 (1986) 1444.\hfill\break

[56] See E.Giannetto, G.D.Maccarrone, R.Mignani and E.Recami:
 {\em Phys. Lett.}  B178 (1986) 115-120, and
refs. therein.\hfill\break

[57] See e.g. R.G.H.Robertson et al.:  {\em Phys. Rev.
Lett.}  67 (1991) 957; A.Burrows et al.:
 {\em Phys. Rev. Lett.}  68 (1992) 3834;
Ch.Weinheimer et al.:  {\em Phys. Lett.}
B300 (1993) 210; E.Holtzshuh et al.:  {\em Phys.
Lett.}  B287 (1992) 381; H.Kawakami et al.:
 {\em Phys. Lett.}  B256 (1991) 105, and so
on. See also the reviews and comments by M.Baldo Ceolin: ``Review
of neutrino physics", invited talk at the  {\em XXIII
Int. Symp. on Multiparticle Dynamics}  (Aspen, CO;
Sept.1993); E.W.Otten:  {\em Nucl. Phys.
News}  5 (1995) 11.\hfill\break

[58] J.A.Zensus and S.Unwin (editors):  {\em Superluminal
Radio Sources}  (Cambridge Univ.Press; Cambridge, 1987),
and references therein.\hfill\break

[59] I.F.Mirabel and L.F.Rodriguez: {\em Nature}  371 (1994) 46
[with a  {\em Nature'} s comment, ``A
galactic speed record", by G.Gisler, at page 18 of the same
issue]; \ S.J.Tingay et al.: {\em Nature}  374 (1995) 141.\hfill\break

[60] E.Recami, A.Castellino, G.D.Maccarrone and M.Rodon\`{o}: {\em Nuovo
Cimento}  B93 (1986) 119.\hfill\break

[61]  {\em Scientific American:}  an article
in the Aug. 1993 issue;  {\em Nature:}
comment ``Light faster than light?" by R.Landauer, Oct. 21,
1993;  {\em New Scientist:}  editorial
``Faster than Einstein" at p.3, plus an article by J.Brown at
p.26, April 1995.\hfill\break

[62] Cf. V.S.Olkhovsky and E.Recami:  {\em Phys.
Reports}  214 (1992) 339, and refs. therein.\hfill\break

[63] A.Enders and G.Nimtz:  {\em J. de
Physique-I}  2 (1992) 1693; 3 (1993) 1089;
 {\em Phys. Rev.}  B47 (1993) 9605;
 {\em Phys. Rev.}  E48 (1993) 632; G.Nimtz,
A.Enders and H.Spieker:  {\em J. de
Physique-I}  4 (1994) 1; W.Heitmann and G.Nimtz:
 {\em Phys. Lett.}  A196 (1994) 154; G.Nimtz,
A.Enders and H.Spieker: in  {\em Wave and Particle in
Light and Matter}  (Proceedings of the Trani Workshop,
Italy, Sept.1992), A.van der Merwe and A.Garuccio editors. New York:
Plenum, in press; H.Aichmann and G.Nimtz, ``Tunnelling of a
FM-Signal: Mozart 40," submitted for pub.; G.Nimtz and W.Heitmann:
{\em Prog. Quant. Electr.} 21 (1997) 81-108.\hfill\break

[64] A.M.Steinberg, P.G.Kwiat and R.Y.Chiao:  {\em Phys.
Rev. Lett.}  71 (1993) 708; R.Y.Chiao, P.G.Kwiat and
A.M.Steinberg:  {\em Scientific American}
269 (1993), issue no.2, p.38.  See also A.M.Steinberg and R.Y. Chiao:
Phys. Rev. A51 (1995) 3525; P.G.Kwiat, A.M.Steinberg, R.Y.Chiao,
P.H.Eberhard and M.D.Petroff: Phys. Rev. A{\bf 48} (1993) R867; \ E.L.Bolda,
R.Y.Chiao and J.C.Garrison: Phys. Rev. A{\bf 48} (1993) 3890; \
A.M.Steinberg: {\em Phys. Rev. Lett.} 74 (1995) 2405.\hfill\break

[65] A.Ranfagni, P.Fabeni, G.P.Pazzi and D.Mugnai:
 {\em Phys. Rev.}  E48 (1993) 1453.\hfill\break

[66] Ch.Spielmann, R.Szipocs, A.Stingl and F.Krausz:  {\em Phys.
Rev. Lett.}  73 (1994) 2308.\hfill\break

[67] V.S.Olkhovsky, E.Recami, F.Raciti and A.K.Zaichenko:
 {\em J. de Physique-I}  5 (1995) 1351-1365.
See also pages 158 and 116-117 of ref.[7]; \ D.Mugnai et al.:
{\em Phys. Lett.}  A209 (1995) 227-234; \ G.Privitera: ``Tempi di
Tunnelling", MsSc thesis supervised by E.Recami (Catania Univ.; 1995).

\end{document}